\documentclass[11pt,letterpaper]{article}

\pdfoutput=1
\usepackage{times}
\usepackage{fancyhdr}
\usepackage[sf,bf,compact]{titlesec}
\usepackage{pdfpages}
\usepackage{wrapfig}
\usepackage[margin=1in]{geometry}
\usepackage{xcolor}
\usepackage{acronym}

\usepackage[labelfont=bf,labelsep=space,list=true]{subcaption}

\usepackage{authblk}
\usepackage{microtype}
\usepackage{tikz}
\usetikzlibrary{patterns, arrows, backgrounds, shapes, positioning, shadows, er,
  decorations.pathmorphing}

\pdfpageattr{/Group <</S /Transparency /I true /CS /DeviceRGB>>}

\RequirePackage{color}\definecolor{RED}{rgb}{1,0,0}\definecolor{BLUE}{rgb}{0,0,1}\definecolor{GREEN}{rgb}{0,1,0} 
\colorlet{paperblue}{-red!75!green!50}

\let\oldmaketitle\maketitle
\renewcommand\maketitle{{\bfseries\sffamily\oldmaketitle}}

\title{A Remote Procedure Call Approach for Extreme-scale Services}

\author[1]{Jerome Soumagne}
\author[2]{Philip H. Carns}
\author[3]{Dries Kimpe}
\author[1]{Quincey Koziol}
\author[2]{Robert B. Ross}

\affil[1]{The HDF Group}
\affil[2]{Argonne National Laboratory}
\affil[3]{KCG}

\date{}

\begin{document}

\maketitle

\vspace{-40pt}

\thispagestyle{empty}
\pagenumbering{gobble}


\section*{Introduction}

When working at exascale, the various constraints imposed by the extreme
scale of the system bring new challenges for application users and 
software/middleware developers. In that context, and to provide best performance,
resiliency and energy efficiency, software may be provided as a
service oriented approach, adjusting resource utilization to best meet facility
and user requirements. These services, which can offer various
capabilities, may be used on demand by a broad range of applications.

Remote procedure call (RPC)~\cite{Birrell1984} is a technique that originally
followed a client/server model and allowed local calls to be
transparently executed on remote resources. RPC consists of serializing the local
function parameters into a memory buffer and sending that buffer to a remote
target that in turn deserializes the parameters and executes the corresponding
function call, returning the result back to the caller. Building reusable services 
requires the definition of a communication model to remotely access these services
and for this purpose, RPC can serve as a foundation for accessing them.
We introduce the necessary building blocks to enable this ecosystem to software and
middleware developers with an RPC framework called \textit{Mercury}~\cite{Soumagne2013}.

\section*{RPC for High-Performance Computing}
\begin{figure}[!htb]
\centering
\tikzstyle{endBox} = [rectangle, rounded corners, minimum height=60pt, text centered,
  text width=40pt, font=\small \bfseries, drop shadow={opacity=0.25}]
\tikzstyle{subBox} = [rectangle, draw, thick, dashed, rounded corners,
  fill=white, minimum height=20pt, text centered, font=\scriptsize \bfseries]
\tikzstyle{trans} = [->, >=stealth, thick, text centered, font=\footnotesize]
\tikzstyle{message} = [draw=black, ->, thick, text centered, text width=100pt,
  font=\footnotesize]

\begin{tikzpicture}
\node[endBox, draw, text=white, fill=paperblue] (client) {Origin};
\node[right=120pt of client.east] (client_east) {};
\node[endBox, draw, text=white, fill=paperblue, right=45pt of client_east] (server) {Target};

\node[below=10pt of client.north east] (client_north1) {};
\node[subBox] at (client_north1) (serialize) {RPC proc};

\node[right=25pt of client.north east] (client_north) {};
\node[above=2pt of client_north] (client_extra_north) {};

\node[above=10pt of client.south east] (client_south1) {};
\node[right=25pt of client.south east] (client_south) {};
\node[below=2pt of client_south] (client_extra_south) {};

\draw[paperblue, decorate, decoration=zigzag, thick] (client_extra_north) --
  node [text=paperblue, font=\footnotesize \itshape, below=40pt, text centered]
  {Network Abstraction Layer}
  (client_extra_south);

\node[below=10pt of server.north west] (server_north1) {};
\node[subBox] at (server_north1) (deserialize) {RPC proc};

\node[left=25pt of server.north west] (server_north) {};
\node[above=2pt of server_north] (server_extra_north) {};

\node[above=10pt of server.south west] (server_south1) {};
\node[left=25pt of server.south west] (server_south) {};
\node[below=2pt of server_south] (server_extra_south) {};

\draw[paperblue, decorate, decoration=zigzag, thick] (server_extra_north) --
  node [] {} (server_extra_south);

\draw[message] (serialize) to node [above] {Metadata\\ (Point-to-point messaging)} (deserialize);
\draw[message] (client_south1) to node [above] (bk) {Bulk Data\\ (RDMA transfer)} (server_south1);

\end{tikzpicture}
\vspace{-10pt}
\caption{Architecture overview.}
\label{fig:overview}
\end{figure}
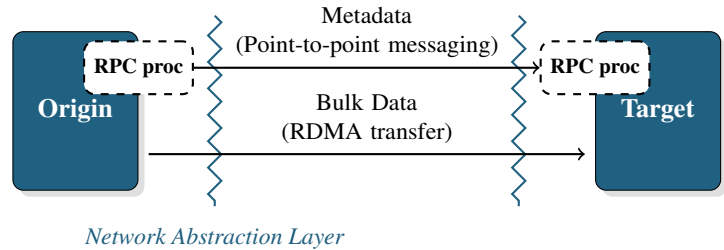

RPC appears to be useful as a basis for building services for high-performance computing.
However, using standard and generic RPC frameworks on a high-performance
computing (HPC) system presents two main limitations: the inability to take
advantage of the native high-speed transport mechanism in order to transfer data
efficiently, since RPC frameworks are mainly designed on top of
TCP/IP protocols, and the inability to transfer very large amounts of data, since
the limit imposed by common RPC interfaces is generally on the order of a
megabyte. In addition, even if no size limit is enforced, transferring large
amounts of data through an RPC library is usually discouraged, mostly because of
overhead from serialization and encoding, causing the data to be copied
many times before reaching the remote node. Mercury is designed to address these 
limitations by taking advantage of native high-speed interconnects and exposing
the semantics required for making nonblocking RPC as
well as for supporting large data arguments, represented in
figure~\ref{fig:overview}.

\section*{Basis for Reusable Services}
To serve as a basis for accessing and enabling reusable services in
a high-performance computing environment, Mercury is designed to be both
easily integrated and extended as well as providing a model that enables both
high-performance and high-concurrency.
It provides a network plugin mechanism that can support existing as well as future
network fabrics, abstracted by a network abstraction layer.
This network abstraction layer provides only the minimal necessary
set of functionality and therefore makes it easy for developers
to create a new plugin. Mercury builds upon this layer and through it defines
an RPC operation as a lightweight operation, which consists of a buffer transmitted
to a target where a function callback is executed. Serialization and
deserialization of arguments can be either provided by Mercury or
left to upper layers, which may require more specific encoding/decoding operations.

As services need to interact between each other and coordinate operations in
a dynamic fashion, it is also important for Mercury processes to not be bound
to a specific role, i.e. client or server. Therefore, client and server concepts
are abstracted by the notion of origin and target. An origin process issues
a call to a remote target process. These notions aim at simplifying the semantics
and avoid any real distinction between a client and a server,
since a client may also become a server in the future.

Finally to enable high-concurrency, the Mercury progress and execution model is
based on a callback model, as opposed to a standard request based model. When a
Mercury operation completes, a user-provided function callback is placed onto
a completion queue before it gets executed. This
has two advantages: first it allows upper layer services to build on top of Mercury
to easily schedule operations by using for instance, a multithreaded execution
model; second, it still allows definition when necessary and more convenient of
shim layers that simplify common cases, based for instance on a request model to
provide post/test operations.

\section*{Conclusion}
Defining reusable software services at exascale is an upcoming challenge.
As such Mercury will be a valuable asset and serve as a basis by providing a
lightweight and modular RPC infrastructure for high-performance computing
middleware, enabling both high-speed transfers and high-concurrency.

Higher-level features such as multithreaded execution, pipelining operations,
or other auxiliary features such as group membership, authorization, etc, are
not provided by Mercury directly, although Mercury is designed to provide the
ecosystem so that these features can easily be built on top of it.

\bibliographystyle{abbrv}
\bibliography{main}

\end{document}